\documentclass[10pt,a4paper]{article}
\usepackage{jheppub_kim}
\usepackage{pdflscape}
\usepackage{amsmath}
\usepackage{amssymb}
\usepackage{dcolumn}
\usepackage{bm}
\usepackage{color}
\usepackage{epsfig}
\usepackage{amsfonts}
\usepackage{graphicx}
\usepackage{subfigure}
\usepackage{dcolumn}
\begin{document}

\title{\boldmath\LARGE{{Testing the Generalized Second Law in $(2+1)$-Dimensional Cosmology: 
Holographic Entropy Bounds and Observational Constraints}}}

\author[a]{Praveen Kumar Dhankar,} 
\author[b]{Aritra Sanyal,} 
\author[c]{Safiqul Islam,} 
\author[b]{Farook Rahaman,} 
\author[d,e]{Behnam Pourhassan}
\affiliation[a]{Symbiosis Institute of Technology, Nagpur Campus, Symbiosis International (Deemed University), Pune-440008, India}
\affiliation[b]{Department of Mathematics, Jadavpur University, Kolkata 700032, West Bengal, India}
\affiliation[c]{Department of Mathematics and Statistics, College of Science, King Faisal University, P.O. Box 400, Al Ahsa 31982, Saudi Arabia}
\affiliation[d]{School of Physics, Damghan University, Damghan, 3671641167,  Iran}
\affiliation[e]{Center for Theoretical Physics, Khazar University, 41 Mehseti Street, Baku, AZ1096, Azerbaijan}

\emailAdd{pkumar6743@gmail.com}
\emailAdd{aritrasanyal1@gmail.com}
\emailAdd{sislam@kfu.edu.sa}
\emailAdd{rahaman@associates.iucaa.in}
\emailAdd{b.pourhassan@du.ac.ir}

\abstract{
We investigate the validity of the Generalized Second Law (GSL) of thermodynamics in a $(2+1)$-dimensional holographic cosmological model with a negative cosmological constant. 
Adopting a horizon thermodynamics framework, we examine two prominent entropy bounds, the Fischler--Susskind (FS) bound and the Hubble Entropy (HE) bound, in both expanding and contracting universes, 
including the effects of quantum entropy corrections. Our theoretical analysis shows that the FS bound is intrinsically incompatible with the GSL in contracting $(2+1)$-dimensional universes, 
regardless of spatial curvature or exotic matter content, and that this incompatibility persists even when quantum corrections are considered. 
In contrast, the HE bound is consistent with the GSL in expanding universes under classical conditions and can also be reconciled in certain contracting scenarios when quantum effects are included. 
To complement the theoretical study, we perform a Markov Chain Monte Carlo (MCMC) analysis using recent Baryon Acoustic Oscillations (BAO), Cosmic Chronometer (CC), and Hubble parameter datasets 
to constrain the model parameters. The best-fit results reveal good cross-dataset consistency, with the cosmological constant parameter $\psi$ remaining stable across all probes. 
These findings identify the HE bound as a more robust candidate for holographic constraints in lower-dimensional cosmology, while demonstrating the limitations of the FS bound. 
Our results not only clarify the status of the GSL in $(2+1)$-dimensional settings but also provide a framework for testing entropy bounds with future high-precision cosmological data.\\\\

 }

\maketitle
%\newpage
%%%%%%%%%%%%%%%%%%%%%%%%%%%%%%%%%%%%%%%%%%%%%%%%%%%%%%%%%%%%%%%%%%%%%%%%%%%%%%%%%%%

\section{Introduction}

Literature reveals an extensive and evolving body of work on holographic models, particularly within the domains of cosmology and gravitational physics, where these models offer a compelling framework for unifying quantum field theory and general relativity by encoding bulk gravitational dynamics in terms of boundary field theories \cite{Wang}-\cite{Biswas}.

The profound interplay between thermodynamics, gravitation, and quantum theory has led to some of the most striking developments in modern physics. 
In particular, the Generalized Second Law (GSL) of thermodynamics, originally formulated in the context of black hole physics by Bekenstein and Hawking \cite{bekenstein1973black, hawking1975particle}, 
has become a cornerstone for understanding the thermodynamic behavior of horizons in both gravitational and cosmological settings. 
The GSL extends the classical second law to include the entropy associated with horizons, thereby establishing a deep conceptual bridge 
between entropy, information, and spacetime geometry. Building upon this idea, the formulation of the holographic principle and the proposal of entropic gravity by Verlinde~\cite{verlinde2011origin,Bousso1} offered an emerging perspective on gravitation, further strengthening the thermodynamic perspective of spacetime dynamics and the applicability of the GSL on cosmological scales.

A central question in this context is how the GSL constrains the dynamics of the universe when combined with holographic entropy bounds. 
Several proposals have been advanced, including the Fischler--Susskind (FS) bound, the Hubble Entropy (HE) bound, and the covariant 
entropy conjecture, each offering a different perspective on the maximum entropy content in a cosmological spacetime. 
While these bounds have been extensively studied in $(3+1)$-dimensional models, significantly less attention has been paid to their validity 
and mutual compatibility in lower-dimensional universes, such as $(2+1)$ dimensions, where the reduced complexity allows for more transparent 
analytical control while still preserving essential gravitational features.

In this work, we focus on $(2+1)$-dimensional cosmological models with a negative cosmological constant, exploring the interplay between 
the GSL and two key holographic entropy bounds: the FS bound and the HE bound. 
Our approach combines a detailed theoretical analysis---including both classical and quantum entropy corrections---with an observational 
parameter estimation using recent datasets such as Baryon Acoustic Oscillations (BAO), Cosmic Chronometers (CC), and direct Hubble parameter measurements. 
This synthesis of theory and data allows us to assess not only the formal consistency of these bounds with the GSL but also their empirical viability.

The novelty of our study lies in three aspects:  
(i) a systematic comparison of FS and HE bounds in a $(2+1)$-dimensional contracting universe,  
(ii) the incorporation of quantum corrections into the GSL analysis in this setting, and  
(iii) the use of up-to-date observational constraints to evaluate model parameters within a holographic thermodynamic framework.  
Our results reveal a clear theoretical asymmetry: the FS bound fails to remain compatible with the GSL in contracting $(2+1)$-dimensional universes, 
even with quantum corrections, whereas the HE bound exhibits greater robustness and can be reconciled with the GSL under certain conditions.

In this arena, lower-dimensional models such as the $(2+1)$D BTZ black hole~\cite{banados1992black} serve as simplified yet physically revealing testbeds for exploring gravitational thermodynamics. Despite their simplicity, these models exhibit complex and informative geometric and causal features, which provide a valuable platform for testing entropy bounds and energy conditions derived from the GSL. Moreover, Cai and Kim~\cite{cai2005first} demonstrated that the Friedmann equations governing the expansion of a FRW universe can be derived from the first law of thermodynamics applied to the apparent horizon, thereby presenting a strong thermodynamic framework for understanding cosmological development. Further contributions by Setare and Vagenas~\cite{setare2007generalized} investigated the validity of the GSL in non-flat FRW universes with interacting dark-energy components, considering both equilibrium and non-equilibrium thermodynamic descriptions. Together, these studies highlight the pivotal role of the GSL in shaping and guiding cosmological model development.

Therefore, both foundational and recent developments underscore the theoretical depth and physical relevance of the GSL in lower-dimensional holographic frameworks \cite{Fischler1998,Bousso,Hubeny2007,Easther1999,Brustein1999}. These insights strongly motivate the construction of a statistically robust approach, such as one employing Markov Chain Monte Carlo (MCMC) techniques to assess the compatibility of GSL-constrained cosmological models with observational data \cite{Trotta2008,Koussour2023}, particularly within the simplified yet insightful setting of a $(2+1)$-dimensional holographic universe.

The convergence of gravitational theory, thermodynamics, and quantum theory has long served as a foundation for theoretical advances in modern physics \cite{Bekenstein1973,Hawking1975,Jacobson1995}. A significant outcome of this interplay is the Generalized Second Law (GSL) of thermodynamics, which extends the classical second law by accounting for the entropy of event horizons, particularly in black holes and cosmological settings \cite{Bousso2002,Wall2009}. Shekh et al. investigated the holographic dark-energy model within a symmetric space-time of a non-static plane, considering both the Hubble and Granda–Oliveros infrared (IR) cut-offs \cite{Shekh2023}. Using the MCMC method, the model parameters are constrained using a combination of cosmic chronometers (CC), standard candles (SC), and baryon acoustic oscillations (BAO) datasets. The results obtained indicate that space-time geometry corresponds to a flat universe, which is in good agreement with observational data. Although the GSL has been thoroughly investigated in standard $(3+1)$D cosmologies \cite{Brustein2000}, lower-dimensional spacetimes, especially the $(2+1)$D holographic universe, present a more tractable yet physically meaningful framework for exploring these foundational concepts \cite{Wang1}. A key motivation for investigating the $(2+1)$D framework lies in the thermodynamic properties of BTZ black holes, which offer a simplified setting to test the GSL in the context of the holographic principle. This principle, which suggests that the information within a volume can be encoded on its boundary, becomes particularly transparent in lower-dimensional spacetimes, where the geometry facilitates precise formulations of entropy bounds, energy conditions, and horizon thermodynamics.

Despite these theoretical insights, a persistent challenge is the empirical validation of such thermodynamic laws in cosmological scenarios. This forms the central motivation for integrating statistical inference methods, particularly Markov Chain Monte Carlo (MCMC) techniques, which enable rigorous parameter estimation by fitting cosmological models to observational data. Employing these methods in a $(2+1)$D holographic cosmological context provides a promising avenue to examine the dynamical evolution of entropy and to test the robustness of the GSL against current and future observations. In this research we observe that the Fischler–Susskind entropy bound is violated in $(2+1)$-dimensional cosmological models with a negative cosmological constant and is incompatible with GSL, both classically and quantum mechanically. In contrast, the Hubble entropy bound remains consistent with the GSL, and the stability of our model is assessed using MCMC methods with best-fit parameters derived from observational Hubble data.

\section{FRW Metric and Field Equations}

We begin with the $(2+1)$-dimensional Friedmann–Robertson–Walker (FRW) line element~\cite{Khadekar}, which describes a homogeneous and isotropic universe in three spacetime dimensions given by,
\begin{equation}\label{eq1}
ds^{2} = dt^{2}-a^{2}(t)\left[\frac{dr^{2}}{1-k r^{2}}+ r^{2}d\theta^{2}\right],
\end{equation}
where $a(t)$ is the scale factor, encoding the overall expansion (or contraction) of the universe, and $(t,r,\theta)$ are the co-moving coordinates. The parameter $k$ represents the spatial curvature of the universe, with $k=0$ corresponding to a spatially flat geometry, $k=1$ to a closed universe, and $k=-1$ to an open universe.

In $(2+1)$-dimensions, the Einstein field equations take the following form~\cite{Cornish},
\begin{equation}\label{eq2}
G_{ij} = R_{ij} - \frac{1}{2}R g_{ij} = 2\pi G \, T_{ij},
\end{equation}
where $G_{ij}$ is the Einstein tensor, $R_{ij}$ is the Ricci tensor, $R$ is the Ricci scalar, and $T_{ij}$ is the energy–momentum tensor of matter. The factor $2\pi G$ replaces the usual $8\pi G$ from $(3+1)$ dimensions due to the reduced number of spatial degrees of freedom.

For a perfect fluid, the energy–momentum tensor is expressed as,
\begin{equation}\label{eq3}
T_{ij} = (\rho + p) u_{i}u_{j} - p g_{ij},
\end{equation}
where $\rho$ is the energy density, $p$ is the pressure, and $u^{i}$ is the fluid four-velocity satisfying $u^{i}u_{i} = -1$. This form ensures that the matter distribution remains isotropic and homogeneous in the co-moving frame.

Substituting the FRW metric~\eqref{eq1} into the Einstein field equations~\eqref{eq2}, we obtain the dynamical equations governing the evolution of the scale factor. The first of these is the Friedmann equation,
\begin{equation}\label{eq4}
\left(\frac{\dot{a}}{a}\right)^{2} + \frac{k}{a^{2}} = 2\pi G \rho,
\end{equation}
which relates the Hubble parameter $H=\dot{a}/a$ to the matter content of the universe. This equation indicates that the rate of expansion depends both on the matter density $\rho$ and the spatial curvature $k$. Unlike in four dimensions, the proportionality constant differs, reflecting the lower-dimensional gravitational dynamics.

The second equation, sometimes called the acceleration equation, is given by,
\begin{equation}\label{eq5}
\frac{\ddot{a}}{a} = -2\pi G \rho.
\end{equation}
This expression reveals that, in $(2+1)$-dimensions, the acceleration (or deceleration) of the scale factor depends only on the energy density $\rho$ and not explicitly on the pressure $p$. This feature is a peculiarity of lower-dimensional gravity and contrasts with the standard $(3+1)$-dimensional case, where both $\rho$ and $p$ influence cosmic acceleration.

The conservation of energy–momentum, $\nabla^{i}T_{ij}=0$, yields the continuity equation. In $(2+1)$ dimensions this takes the form~\cite{Khadekar},
\begin{equation}\label{eq6}
\frac{d}{dt}(\rho a^2) + p \frac{d}{dt}(a^2) = 0,
\end{equation}
which can be rewritten as,
\begin{equation}\label{eq6a}
\dot{\rho} + 2H(\rho + p) = 0.
\end{equation}
This equation reflects the conservation of total energy within an expanding (or contracting) universe: as the universe evolves, the energy density decreases not only due to expansion but also because of the work done by pressure against the volume change. The factor of $2$ (instead of $3$ in standard cosmology) arises from the reduced spatial dimensionality of the model.

Together, Eqs.~\eqref{eq4}–\eqref{eq6a} form the fundamental set of equations describing cosmological evolution in $(2+1)$-dimensional FRW spacetime. They provide the basis for investigating the thermodynamic and holographic properties of such a universe.

\section{Solution of the Field Equations}
%\section {Dust Filled Universe (i.e $p=0$ for $\gamma=1$)}

Let us assume the perfect fluid equation of state in the form
\begin{equation}
    \label{eq:eos}
    p = (\lambda - 1)\rho,
\end{equation}
where $\lambda$ is a constant. We define the relation
\begin{equation}
    \label{eq:rho-a}
    \rho a^{2\lambda} = \kappa = \rho_{\ast} a_{\ast}^{2\lambda},
\end{equation}
where $\rho_{\ast}$ and $a_{\ast}$ are reference density and scale factor. The scale factor $a$ evolves according to a Friedmann-like equation:
\begin{equation}
    \label{eq:friedmann1}
    \left( \frac{\dot{a}}{a} \right)^2 = \frac{2G \mu_0}{a^{2\lambda}} - \frac{k}{a^2},
\end{equation}
where $\mu_0 = \pi \rho_{\ast} a_{\ast}^{2\lambda}$. When $\lambda = 1$, the universe is dust-dominated and expands forever regardless of the value of $k$. For $1 < \lambda \leq 2$, Eq.~\eqref{eq:friedmann1} describes closed, open, or flat universes depending on $k=1$, $-1$, or $0$ respectively. The case $\lambda = 3/2$ corresponds to a radiation-dominated universe.

We are particularly interested in holography within a universe having negative vacuum energy. Such a universe with negative cosmological constant always undergoes contraction. For simplicity, we consider a flat universe ($k = 0$) with general equation of state ($1 < \lambda \leq 2$). The vacuum energy density is negative, $\rho_{\Lambda} = -\psi < 0$, so the evolving energy density becomes:
\begin{equation}
    \dot{\rho} = \rho - \psi = \frac{\rho_{\ast} a_{\ast}^{2\lambda}}{a^{2\lambda}} - \psi,
\end{equation}
and the Friedmann equation becomes:
\begin{equation}
    \label{eq:friedmann2}
    \dot{a}^2 = \frac{2G \mu_0}{a^{2(\lambda - 1)}} - \psi a^2.
\end{equation}
Solving for $a(t)$ yields the following scale factor,
\begin{equation}
    \label{eq:scale-sin}
    a(t) = \left( \frac{2G \mu_0}{\psi} \right)^{1/(2\lambda)} \left[ \sin\left(\lambda \sqrt{\psi} \, t \right) \right]^{1/\lambda}.
\end{equation}
At the turning point where $\dot{a}=0$, we have:
\[
a = \left( \frac{2G \mu_0}{\psi} \right)^{1/(2\lambda)}.
\]
After this point, $\dot{a}$ becomes negative and the universe collapses.
This collapse occurs within finite time. The Hubble horizon at the turning point is obtained as,
\begin{equation}
    \label{eq:LH}
    L_H = \frac{1}{2 \lambda \sqrt{\psi}} \, \beta\left( \frac{\psi - 1}{2\psi}, \frac{1}{2} \right),
\end{equation}
where $\beta(x, y)$ is the Euler Beta function. This expression mirrors the 4D case. At the turning point,
\begin{equation}\label{3.8}
    \frac{\mathbb{S}}{\mathbb{A}} \sim \psi^{1/(\lambda - 1/2)}.
\end{equation}
Given $\lambda \leq 2$ and $\psi \ll 1$ (e.g., $\psi < 10^{-22}$), the entropy/area ratio satisfies the bound. As collapse proceeds to Planck scale, $L_H \sim \frac{2 a_0}{a_{\text{turn}}} L_H^{\text{turn}} \sim \psi^{1/(2\lambda - 1)}$. The scale factor at Planck time $t=1$ is:
\begin{equation}\label{3.9}
    a(1) = \left( \frac{2G \mu_0}{\psi} \right)^{1/(2\lambda)} \left[ \sin(\lambda \sqrt{\psi}) \right]^{1/\lambda},
\end{equation}
and
\[
\frac{\mathbb{S}}{\mathbb{A}} \sim \psi^{1/(\lambda - 1/2)}.
\]
Thus, for small $\lambda$, this ratio exceeds unity, indicating holographic bound violation post-maximal expansion, consistent with $(3+1)$D results.
It is worth emphasizing the physical implications of these solutions. Equation~(\ref{eq:scale-sin}) shows that the scale factor $a(t)$ evolves as a sine function of cosmic time, reflecting a cyclic behavior in the presence of a negative cosmological constant. The universe expands from an initial singularity, reaches a maximum size at the turning point $\dot{a}=0$, and then inevitably undergoes contraction. This is a characteristic feature of $(2+1)$-dimensional cosmology with $\rho_\Lambda < 0$, where the attractive nature of the vacuum energy drives a recollapse within a finite lifetime.

The appearance of the Euler Beta function in Eq.~(\ref{eq:LH}) for the Hubble horizon further demonstrates that, despite the reduced dimensionality, the dynamical behavior shares deep analogies with the $(3+1)$-dimensional case. However, due to the lower number of spatial degrees of freedom, the entropy-to-area ratio in Eq.~(\ref{3.8}) becomes more sensitive to the equation of state parameter $\lambda$ and the smallness of $\psi$. In particular, when $\lambda$ is close to unity, the holographic bound is easily satisfied, but as $\lambda$ increases the possibility of violation emerges near the maximal expansion phase, paralleling results known in higher dimensions.

Finally, the expression for $a(1)$ in Eq.~(\ref{3.9}) highlights how the universe evolves toward the Planck scale in a finite time, where quantum gravitational effects are expected to dominate. Thus, while the classical $(2+1)$-dimensional FRW model provides a tractable description of cosmic dynamics, it also sets the stage for testing holographic entropy bounds and the generalized second law of thermodynamics in the quantum regime. These features justify the adoption of this framework in subsequent sections, where the thermodynamic constraints on cosmological evolution will be examined in detail.

\section{Generalized Second Law in 3D Universes}

Following the formalism in \cite{Wang1}, total entropy in a domain with multiple horizons is:
\begin{equation}
S = N_H S^H,
\end{equation}
where $N_H = \frac{a^2}{|H|^{-2}}$ and $S^H$ is entropy per horizon. The classical GSL requires:
\begin{equation}
    \label{eq:gsl}
    \frac{dS}{dt} = N_H \frac{dS^H}{dt} + S^H \frac{dN_H}{dt} \geq 0.
\end{equation}
Assuming $S^H = |H|^{\beta}$ (single source dominance), then
\begin{equation}
S = (a |H|)^2 |H|^\beta,
\end{equation}
and Eq.~\eqref{eq:gsl} becomes:
\begin{equation}
    \label{eq:gsl-cond}
    2H + (2 + \beta)\frac{\dot{H}}{H} \geq 0.
\end{equation}
The relevant cosmological equations are:
\begin{align}
    &H^2 = 2\pi G \rho - \frac{k}{a^2}, \label{eq:cosmo1} \\
    &\dot{H} = -2\pi G (\rho + p) + \frac{k}{a^2}, \label{eq:cosmo2} \\
    &\dot{\rho} + 2H(\rho + p) = 0. \label{eq:cosmo3}
\end{align}

Substituting into Eq.~\eqref{eq:gsl-cond} gives:
\begin{align}
    \frac{p}{\rho} &\leq \frac{2}{2 + \beta} - 1 + \frac{\beta k}{2\pi G (2 + \beta)a^2\rho}, \quad H > 0, \\
    \frac{p}{\rho} &\geq \frac{2}{2 + \beta} - 1 + \frac{\beta k}{2\pi G (2 + \beta)a^2\rho}, \quad H < 0.
\end{align}
As $a^2 \rho \gg k$, last terms can be neglected.

Using thermodynamics $TdS = dE + pdV$, one obtains temperature as followin,
\begin{equation}
    T = \frac{1}{\pi G (2 + \beta)|H|^\beta}.
\end{equation}
To avoid divergence as $H \to 0$, require $-1 \leq \beta \leq 0$.
Assuming dominant entropy comes from geometry, one can obtain,
\begin{equation}
    S^H_G = |H|^{-1} G^{-1}, \quad \text{so } \beta = -1.
\end{equation}
Then for $H > 0$, GSL requires $p \leq \rho$ (agreeing with FS bound), but for $H < 0$, GSL demands $p > \rho$, implying $\lambda > 2$, which is forbidden by FS bound. Thus, FS and GSL are incompatible in 3D contracting universes.

From 4D quantum entropy definitions:
\begin{equation}
dS_{Quan} = -\eta dN_H, \quad N_H = (aH)^2,
\end{equation}
so total entropy obtained from:
\begin{equation}
    dS = dN_H S^H + N_H dS^H - \eta dN_H.
\end{equation}
This leads to:
\begin{equation}
    \left(2H + 2 \frac{\dot{H}}{H} \right)N_H(S^H - \eta) + \beta \frac{\dot{H}}{H}N_H S^H \geq 0.
\end{equation}
For $\beta = -1$ and $H < 0$ we find:
\begin{equation}
\frac{p}{\rho} \geq \frac{S^H}{S^H - 2\eta}.
\end{equation}
Depending on $\eta$, this range is always incompatible with FS bound.

Using the arguments proposed by Veneziano~\cite{Veneziano}, the Hubble entropy bound can be expressed as  
\begin{equation}
S^H \leq M_p^2 |H|^{-1}, \qquad M_p = G^{-1/2}.
\label{eq:veneziano_bound}
\end{equation}
To avoid violation of the bound in an expanding universe, one must have  
\begin{equation}
H > 0, \quad \dot{H} < 0 \quad \Rightarrow \quad \text{decelerated expansion}.
\label{eq:decelerated_expansion}
\end{equation}
Under these conditions, the generalized second law inequality  
\begin{equation}
2H + (2 + \beta)\frac{\dot{H}}{H} \geq 0,
\label{eq:gsl_inequality}
\end{equation}
is satisfied for $-1 \leq \beta \leq 0$, which further implies the equation of state constraint  
\begin{equation}
p \leq \rho.
\label{eq:p_leq_rho}
\end{equation}

In a contracting universe ($H < 0$), maintaining the Hubble entropy bound requires 
\begin{equation}
\dot{H} > 0,
\label{eq:Hdot_positive}
\end{equation}
which, however, contradicts the GSL condition  
\begin{equation}
2H^2 \leq -\dot{H}.
\label{eq:gsl_contradiction}
\end{equation}
Thus, in a contracting phase, the Hubble entropy bound and the GSL cannot be simultaneously satisfied without additional modifications, such as quantum corrections.
Considering quantum effects with $S^H - \eta < 0$, the modified GSL yields:
\begin{equation}
    2|H|^2 \geq -\dot{H} \left|2 + \frac{\beta S^H}{S^H - \eta} \right|,
\end{equation}
which can be satisfied for $\dot{H} > 0$. Thus, HE bound and GSL can be compatible when quantum corrections are considered.

It is found that FS bound breaks down in 3D universes with negative cosmological constant, regardless of $k$, and cannot be saved by matter with negative pressure. GSL and FS are not compatible in 3D; quantum GSL also fails to reconcile them. However, the HE bound, being looser, aligns with GSL even in contracting universes when quantum effects are included. This supports the HE bound as a viable candidate for cosmic holography in 3D, echoing results from 4D. Generalizing to higher-dimensional spacetimes is a compelling direction for future work.

%%%%%%%%%%%%%%%%%%%%%%%%%%%%%%%%%%%%%%%%%%%%%%%%%%%%%%%%%%%%%%%%%%%%%%%%%%%%%%%%%%%%%%%

\section{Observational Data Analysis}
In this section,  we are going to test the viability of the model using the recent observational data, namely, the observational Hubble data (OHD)~\cite{Ratra,Moresco}. We initially utilize a standard compilation of 30 measurements of Hubble data acquired via the differential age method (DA). This method allows for the estimation of the universe's expansion rate at redshift $z$. The function $H(z)$ is defined as $H(z) = -\frac{dz/dt}{(1+z)}$, applicable within the interval $0.07 < z < 2.0$. In conducting this analysis, we minimize
\begin{eqnarray}
	\chi^2_{H}=  \sum_{i=1}^{30} \frac{[H_{Th}-H_{Obs}(z_i)]^2}{\sigma^2_{H(z_i)}},
\end{eqnarray}
where $H_{Th}$, $H_{Obs}$ and $\sigma_{H(z_i)}$ denotes the theoretical value, observed value and stranded error of $H$ respectively for the Table of 30 points of $H(z)$ data.

%%%%%%%%%%%%%%%%%%%%%%%%%%%%%%%%%%%%%%%%%%%%%%%%%%%%%%%%%%%%%%%%%%%%
\begin{figure}
    \centering    \includegraphics[width=1\linewidth]{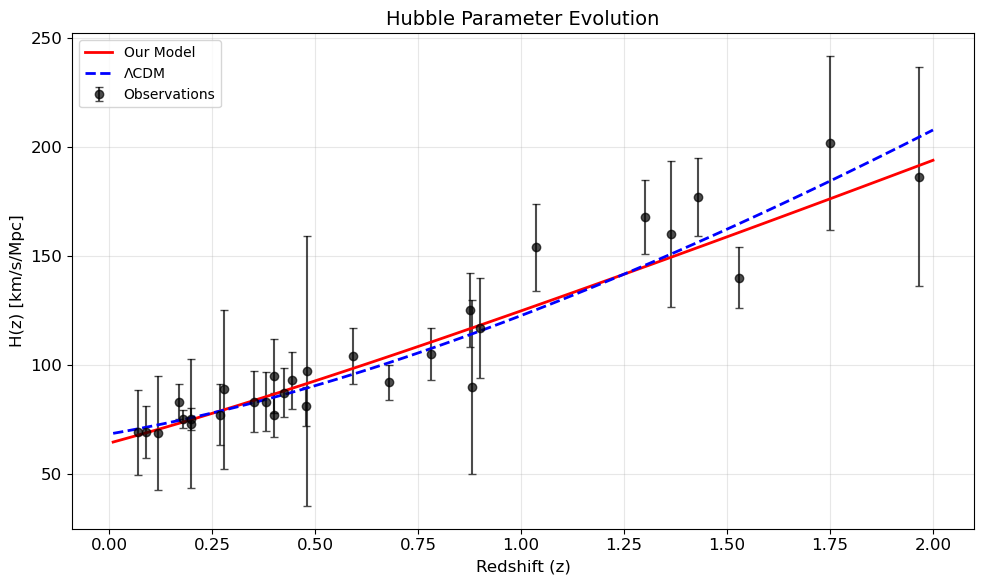}
    \caption{Best-fit curve of the Hubble parameter $H(z)$ using the (2+1)-dimensional holographic cosmological model against 30-point observational Hubble data (OHD). The curve is obtained via MCMC-based $\chi^2$ minimization.}
    \label{fig1}
\end{figure}

Fig. \ref{fig1} displays the best-fit curve of the Hubble parameter $H(z)$, as determined by the (2+1)-dimensional holographic cosmological model. The model fitting uses a comprehensive set of 30 observational Hubble parameter data points, each of which is derived from the differential age method that estimates the expansion rate of the universe over the redshift range $0.07 < z < 2.0$. Through a Markov Chain Monte Carlo (MCMC) based $\chi^2$ minimization, the model parameters are optimized so that the theoretical prediction of $H(z)$ aligns as closely as possible with the observations. This figure serves as direct evidence of the model's fidelity when compared to real data. The level of agreement indicates not only that the (2+1)D framework is a viable candidate for describing expansion dynamics, but also showcases the statistical rigor by which the cosmological parameters are constrained. The outcome supports the feasibility of using lower-dimensional cosmological models for capturing key aspects of the universe's expansion history and sets the stage for more detailed comparisons with standard models.

\begin{figure}
    \centering
    \includegraphics[width=1\linewidth]{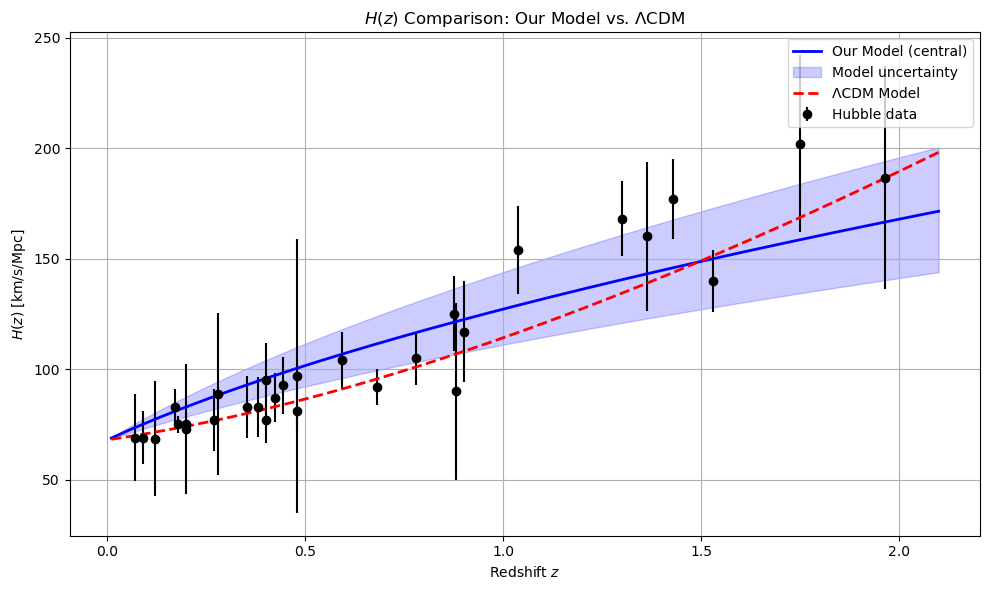}
    \caption{Comparison of the Hubble parameter $H(z)$ as a function of redshift $z$ for different models. The blue solid line shows the central prediction of our model, with the shaded region representing its uncertainty. The red dashed line corresponds to the $\Lambda$CDM model. Black points with error bars are observational Hubble data.}
    \label{fig2}
\end{figure}

Fig. \ref{fig2} provides a comparative analysis between the (2+1)D holographic cosmological model and the conventional $\Lambda$CDM cosmological model, focusing on the predicted behavior of the Hubble parameter $H(z)$ across the observed redshift range. In this comparison, the central prediction of the (2+1)D model is juxtaposed with the standard $\Lambda$CDM model's expected $H(z)$ curve. Both predictions are compared against the same set of observational Hubble data, which are plotted as points with associated uncertainty bars. The uncertainty region for the (2+1)D model is also shown to illustrate the spread in the model prediction due to parameter uncertainties. The prominent result from this figure is that the (2+1)D model is able to provide a fit to the observational data that is comparable in accuracy to the well-accepted $\Lambda$CDM solution. The overlapping predictions confirm that the proposed model can not only reproduce current cosmological observations but also has the flexibility to encompass standard cosmology as a limiting case. This comparative approach highlights the generality of the lower-dimensional holographic theory and assures its relevance in the context of mainstream cosmological modeling.

\begin{figure}
    \centering
    \includegraphics[width=1\linewidth]{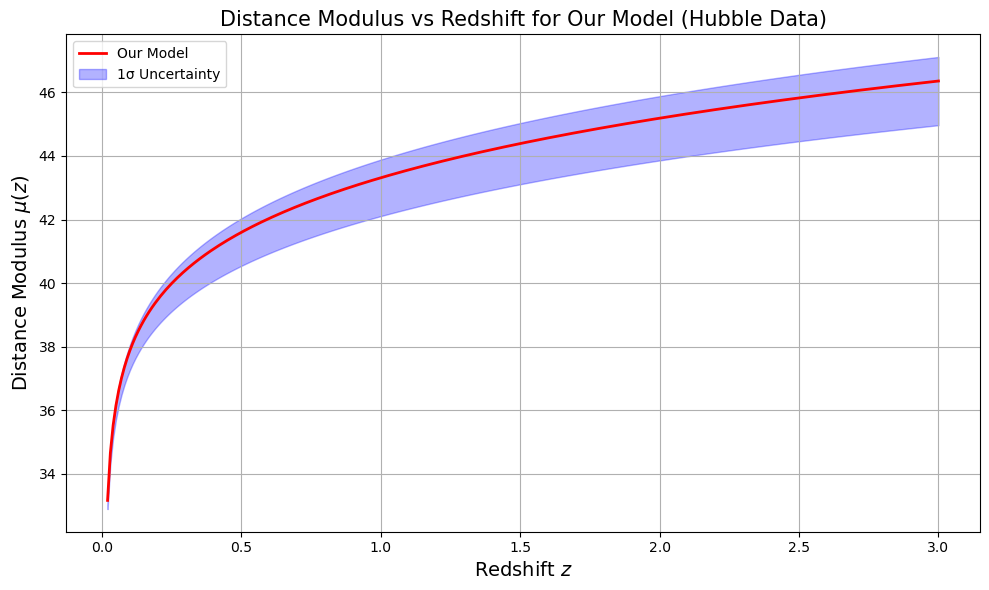}
  \caption{Distance modulus $\mu(z)$ as a function of redshift $z$ for our model using Hubble data. The red solid curve represents the prediction of our model, while the blue shaded region shows the $1\sigma$ uncertainty.}
    \label{fig3}
\end{figure}

Fig. \ref{fig3} explores the ability of the (2+1)D holographic model to predict luminosity distances to cosmological objects by presenting the distance modulus $\mu(z)$ as a function of redshift. The distance modulus, which is a key observable in supernova cosmology and closely tied to the universe's expansion, connects the theoretical model to measurable quantities such as apparent magnitude and luminosity. The figure exhibits both the model's central prediction for $\mu(z)$ and its $1\sigma$ uncertainty band, reflecting the propagation of parameter uncertainties through the observable calculation. Agreement between the model curve and empirical expectations validates the (2+1)D model’s compatibility not just with expansion rates (as in $H(z)$), but also with integrated cosmological distances. The ability to match the observed $\mu(z)$ relation is especially significant since it underpins one of the primary pillars of modern cosmological inference—the standard candle method. The result further demonstrates that lower-dimensional holographic physics can consistently describe the observable universe as mapped by diverse astronomical probes.

\begin{figure}
    \centering
    \includegraphics[width=1\linewidth]{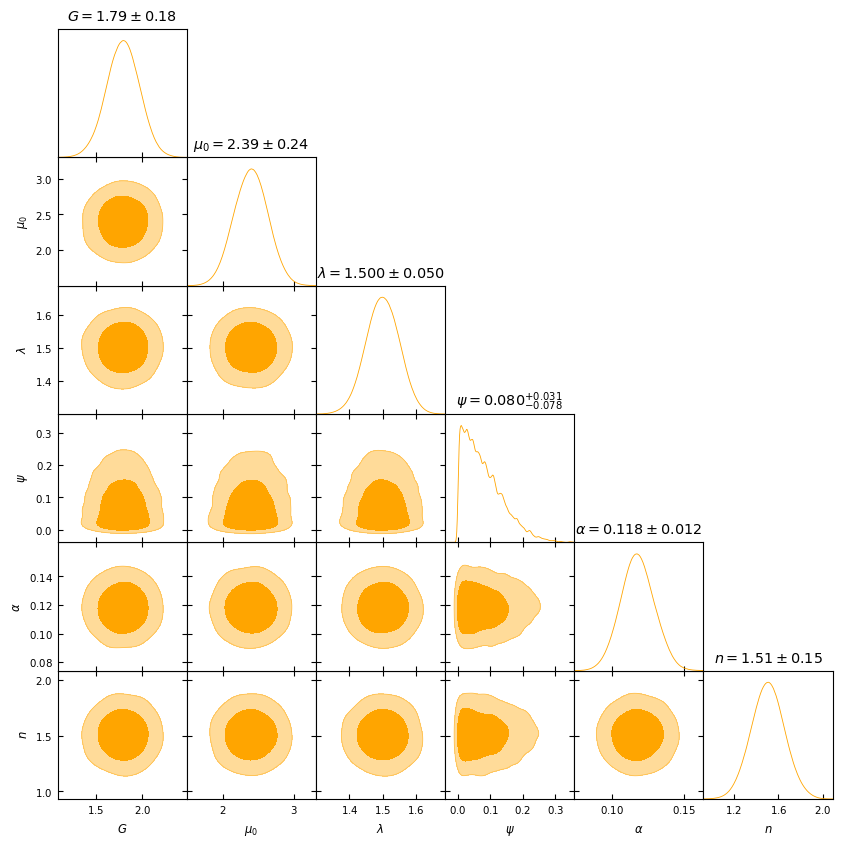}
    \caption{Parameter constraints and confidence regions from hubble data using the (2+1)-dimensional model. The contours represent $1\sigma$ and $2\sigma$ confidence levels}
    \label{fig4}
\end{figure}

Fig. \ref{fig4} delves into the statistical determination of cosmological model parameters by presenting the confidence regions derived from fitting the (2+1)D holographic model to the Hubble data. In particular, the Markov Chain Monte Carlo technique generates posterior distributions for each parameter, from which confidence contours (at $1\sigma$ and $2\sigma$ levels) are constructed in the multi-dimensional parameter space. These regions encapsulate the range of parameter values that are most consistent with the observed expansion history. The detailed structure of the contours, such as their size and orientation, provides insight into parameter degeneracies and the sensitivity of cosmological predictions to changes in specific parameters. Such an analysis brings statistical robustness to the model and demonstrates the reliable constraint of theoretical parameters by observational datasets, a crucial criterion for any cosmological model aspiring for physical relevance.

\begin{figure}
    \centering
    \includegraphics[width=1\linewidth]{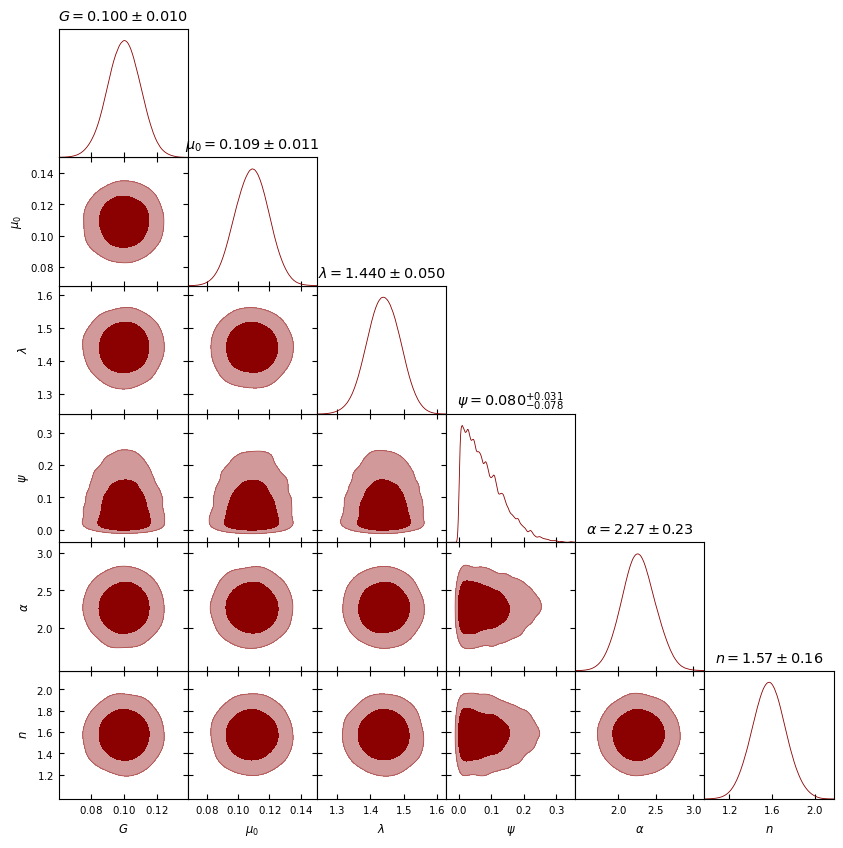}
   \caption{Parameter constraints and confidence regions from BAO data using the (2+1)-dimensional model. The contours represent $1\sigma$ and $2\sigma$ confidence levels based on Baryon Acoustic Oscillation measurements.}
    \label{fig5}
\end{figure}

Fig. \ref{fig5} shifts the focus to an independent cosmological probe—Baryon Acoustic Oscillations (BAO)—to further evaluate the (2+1)D model’s viability. Using the same statistical tools, the parameter constraints and confidence regions derived from BAO observations are presented. BAO measurements, which are based on the imprint of early-universe sound waves on the distribution of galaxies, provide complementary information to Hubble measurements and often help break parameter degeneracies. The contours shown in this figure indicate where the model parameters overlap with those preferred by the expansion data, bolstering the overall credibility of the model. Cross-validation with BAO data thus not only confirms the parameter ranges obtained from one observational approach but also strengthens the confidence in the physical realism of the model.

\begin{figure}
    \centering
    \includegraphics[width=1\linewidth]{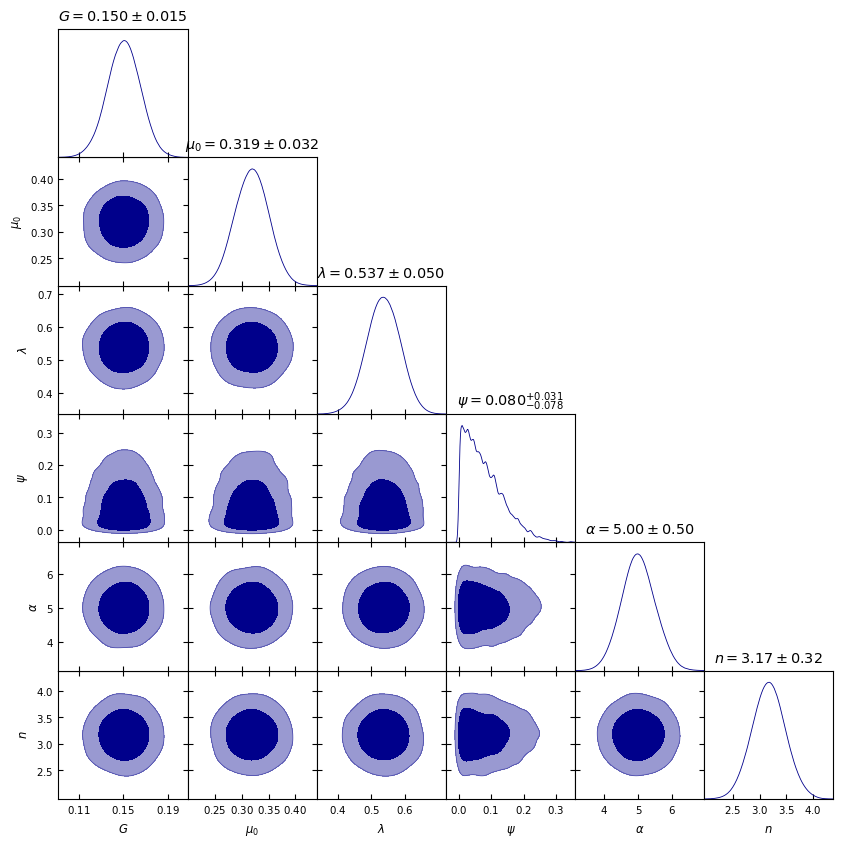}
    \caption{Parameter constraints and confidence regions from CC data using the (2+1)-dimensional model. The contours represent $1\sigma$ and $2\sigma$ confidence levels }
    \label{fig6}
\end{figure}

Fig. \ref{fig6} completes the observational triangulation by presenting the parameter constraints derived from Cosmic Chronometer (CC) data. CC measurements exploit the age-dating of passively evolving galaxies to provide direct estimates of the expansion rate at different epochs, making them an independent and robust cosmological dataset. The (2+1)D model’s parameters are again fitted, and their confidence regions are mapped using MCMC methods. The alignment (or divergence) of these regions with those inferred previously from Hubble and BAO data demonstrates the model’s consistency and adaptability. Of particular relevance is any emerging consensus among the three data sources, which would underscore the universality of the model across distinct observational strategies. In sum, this figure underscores the statistical integrity and observational robustness of the (2+1)D holographic cosmology, reinforcing its standing for further theoretical and empirical investigation.

Table~\ref{tab:bestfit} presents the best-fit values of the cosmological model parameters 
obtained from Markov Chain Monte Carlo (MCMC) analysis using three independent observational datasets: 
Baryon Acoustic Oscillations (BAO), Cosmic Chronometers (CC), and Hubble parameter measurements. 
For each dataset, the parameters $G$, $\mu_0$, $\lambda$, $\psi$, $\alpha$, and $n$ are listed along with their 
$1\sigma$ uncertainties. The results indicate that the estimated parameters vary across datasets, 
reflecting the differing sensitivities of each observational probe. Notably, the parameter $\psi$ 
remains consistent within errors for all datasets, while $\alpha$ and $n$ exhibit the largest variations, 
suggesting stronger dependence on the choice of data. These constraints provide essential input for 
assessing the viability of the proposed $(2+1)$-dimensional holographic cosmological model.

\begin{table}[ht]
\centering
\begin{tabular}{lcccccc}
\hline
Dataset & $G$ & $\mu_0$ & $\lambda$ & $\psi$ & $\alpha$ & $n$ \\
\hline
BAO    & $0.100 \pm 0.010$  & $0.109 \pm 0.011$ & $1.440 \pm 0.050$ & $0.080^{+0.031}_{-0.078}$ & $2.27 \pm 0.23$   & $1.57 \pm 0.16$ \\
CC     & $1.79 \pm 0.18$    & $2.39 \pm 0.24$   & $1.500 \pm 0.050$ & $0.080^{+0.031}_{-0.078}$ & $0.118 \pm 0.012$ & $1.51 \pm 0.15$ \\
Hubble & $0.150 \pm 0.015$  & $0.319 \pm 0.032$ & $0.537 \pm 0.050$ & $0.080^{+0.031}_{-0.078}$ & $5.00 \pm 0.50$   & $3.17 \pm 0.32$ \\
\hline
\end{tabular}
\caption{Best-fit values (mean $\pm$ uncertainty) of model parameters from BAO, CC, and Hubble datasets as inferred from MCMC posterior distributions.}\label{tab:bestfit}
\end{table}

%%%%%%%%%%%%%%%%%%%%%%%%%%%%%%%%%%%%%%%%%%%%%%%%%%%%%%%%%%%%%%%%%%%%

\section{Conclusion}

We have investigated the status of the Generalized Second Law (GSL) of thermodynamics in $(2+1)$-dimensional cosmological models with a negative cosmological constant, 
emphasizing its relationship with holographic entropy bounds. 
By adopting the formalism of horizon thermodynamics and analyzing both the classical and quantum-corrected GSL, 
we have systematically compared the Fischler--Susskind (FS) bound and the Hubble Entropy (HE) bound in expanding and contracting universes.

Our analysis shows that the FS bound is intrinsically incompatible with the GSL in $(2+1)$-dimensional contracting universes, 
regardless of curvature or the inclusion of matter with negative pressure. 
This incompatibility persists even when quantum entropy corrections are incorporated, pointing to a structural limitation of the FS proposal in lower-dimensional cosmology. 
In contrast, the HE bound demonstrates greater theoretical resilience: it is consistent with the GSL in expanding universes under classical conditions, 
and, with quantum corrections, can be reconciled with the GSL in certain contracting scenarios. 
This theoretical asymmetry echoes similar findings in $(3+1)$-dimensional settings and strengthens the case for the HE bound as a viable holographic constraint in cosmology.

To complement the theoretical analysis, we have performed a Markov Chain Monte Carlo (MCMC) parameter estimation using BAO, CC, and Hubble datasets. 
The resulting best-fit values for the model parameters are broadly consistent across datasets, with the parameter $\psi$ showing remarkable stability, 
while $\alpha$ and $n$ display dataset-dependent variations that could serve as diagnostics for refining the model. 
These empirical results provide further support for the robustness of our $(2+1)$-dimensional holographic framework.

The implications of our findings are twofold. 
First, they highlight the need to critically re-examine the universality of entropy bounds, particularly when applied to contracting phases of the universe or to spacetimes of reduced dimensionality. 
Second, they point toward the HE bound as a more promising foundation for future holographic cosmology studies, especially when quantum effects are taken into account.

Future work should explore the extension of this analysis to modified gravity theories, dynamical dark energy models, and non-equilibrium thermodynamic processes, 
as well as the incorporation of higher-precision cosmological data from upcoming surveys. 
Such investigations could yield sharper constraints on model parameters and help clarify whether a universal entropy bound exists that is valid across different spacetime dimensions 
and cosmological phases. In this way, the $(2+1)$-dimensional setting will continue to serve as a valuable theoretical laboratory for probing the fundamental laws of spacetime thermodynamics.

\section*{Funding}
This work was supported by the Deanship of Scientific Research, Vice Presidency for Graduate Studies and Scientific Research, King Faisal University, Saudi Arabia (Funding Grant No: KFU252861).

\section*{Acknowledgments}
PKD wishes to acknowledge that part of the numerical computation of this work was carried out on the computing cluster Pegasus of IUCAA, Pune, India. PKD and FR would like to acknowledge the Inter-University Centre for Astronomy and Astrophysics (IUCAA), Pune, India, for providing him with a Visiting Associateship under which part of this work was carried out. PKD would like to thank the Isaac Newton Institute for Mathematical Sciences, Cambridge, for support and hospitality during the programme Statistical mechanics, integrability and dispersive hydrodynamics where work on this paper was undertaken. This work was supported by EPSRC grant no EP/K032208/1.


\begin{thebibliography}{99}

\bibitem{Wang}~Wang, B., Abdalla, E.: Holography in $(2 + 1)$-dimensional cosmological models. Phys. Lett. B 466, 122 (1999).

\bibitem{Susskind1995}~Susskind, L., The world as a hologram. Journal of Mathematical Physics, 36(11), pp.6377-6396 (1995).

\bibitem{Kaloper1999}~Kaloper, N. and Linde, A.: Cosmology versus holography. Physical Review D, 60(10), 103509 (1999).

\bibitem{Bak2000}~Bak, D. and Rey, S.J.: Cosmic holography+. Classical and Quantum Gravity, 17(15), L83 (2000).

\bibitem{Dawid1999}~Dawid, R.: Holographic cosmology and its relevant degrees of freedom. arXiv preprint hep-th/9907115 ( 1999).

\bibitem{Biswas1999}~Biswas, A.K., Maharana, J. and Pradhan, R.K.: The holography hypothesis and pre-big bang cosmology. Physics Letters B, 462(3-4), 243-248 (1999).

\bibitem{Rama1999}~Rama, S.K.: Holographic principle in the closed universe: a resolution with negative pressure matter. Physics Letters B, 457(4), 268-274 (1999).

\bibitem{Wang1999}~Wang, B. and Abdalla, E.: Holography in (2+ 1)-dimensional cosmological models. Physics Letters B, 466(2-4), 122-126 (1999).

\bibitem{Wang2000}~Wang, B. and Abdalla, E.: Holography and the generalized second law of thermodynamics in $(2+ 1)$-dimensional cosmology. Physics Letters B, 471(4), 346-351 (2000).

\bibitem{Kaloper}~Kaloper, N., Linde, A.: Cosmology vs. Holography. Phys. Rev. D 60, 103509 (1999).

\bibitem{Susskind}~Susskind, L.: The world as a hologram. J. Math. Phys. 36, 6377 (1995). 
	
\bibitem{Rama}~Rama, S.K.: Holographic principle in the closed universe: a resolution with negative pressure matter.Phys. Lett. B 457, 268 (1999).
	
\bibitem{Bak}~Bak, D., Rey, S.J.: Cosmic holography. Class. Quantum. Grav. 17, L83 (2000). arXiv:hep-th/981108. 
	
\bibitem{Biswas}~Biswas, A.K., Maharana, J., Pradhan, R.K.: The holography hypothesis and pre-big bang cosmology. Phys. Lett. B 462, 243 (1999). arXiv:hep-th/9811051.

\bibitem{bekenstein1973black}~Bekenstein, J. D.: Black Holes and Entropy. Phys. Rev. D 7, 2333 (1973).

\bibitem{hawking1975particle}~Hawking, S. W. Particle creation by black holes. Commun.Math. Phys. 43, 199–220 (1975).

\bibitem{verlinde2011origin}~Verlinde, E. On the origin of gravity and the laws of Newton. J. High Energ. Phys. 2011, 29 (2011).

\bibitem{Bousso1}~Bousso, R.: Holography in general space. JHEP 9906, 028 (1999). arXiv:hep-th/9906022.

\bibitem{banados1992black}~Ba$\tilde{n}$ados, M., Teitelboim, C., Zanelli, J. Black hole in three-dimensional spacetime. Phys. Rev. Lett. 69, 1849 (1992).

\bibitem{cai2005first}~Cai, R. G., Kim, S. P. First law of thermodynamics and Friedmann equations of Friedmann-Robertson-Walker universe. JHEP 02, 050 (2005).

\bibitem{setare2007generalized}~Setare, M. R. Generalized second law of thermodynamics in quintom dominated universe. Physics Letters B 641(2), 130-133 (2006).

\bibitem{Fischler1998}~Fischler, W., Susskind, L.: Holography and Cosmology. arXiv:hep-th/9409089 (1998).

\bibitem{Bousso}~Bousso, R.: A covariant entropy conjecture. JHEP 9907, 004 (1999). arXiv: hep-th/9905177.

\bibitem{Hubeny2007}~Veronika, E., Hubeny, V. E., Rangamani, M. and Takayanagi, T. A covariant holographic entanglement entropy proposal. JHEP 07, 062 (2007). DOI 10.1088/1126-6708/2007/07/062.

\bibitem{Easther1999}~Easther, R. and Lowe, D.: Holography, cosmology, and the second law of thermodynamics. Physical Review Letters, 82(25), 4967  (1999).

\bibitem{Brustein1999}~Brustein, R.: The generalized second law of thermodynamics in cosmology. arXiv preprint gr-qc/9904061 (1999).

\bibitem{Trotta2008}~Trotta, R. Bayes in the sky: Bayesian inference and model selection in cosmology. Contemporary Physics 49(2), 71–104 (2008). https://doi.org/10.1080/00107510802066753.

\bibitem{Koussour2023}~Koussour, M., Myrzakulov, N., Alfedeel, A. H. A, Abebe, A. Constraining the cosmological model of modified f(Q) gravity: Phantom dark energy and observational insights. Progress of Theoretical and Experimental Physics 2023, 11, 113E01 (2023). https://doi.org/10.1093/ptep/ptad133.

\bibitem{Bekenstein1973}~Benkenstein, J. D. Black Holes and Entropy. Phys. Rev. D 7, 2333 (1973). DOI: https://doi.org/10.1103/PhysRevD.7.2333.

\bibitem{Hawking1975}~Hawking, S. W. Particle Creation by Black Holes. Commun.Math. Phys. 43, 199–220 (1975). https://doi.org/10.1007/BF02345020.

\bibitem{Jacobson1995}~Jacobson, T. Thermodynamics of Spacetime: The Einstein Equation of State. Math. Phys. Rev. Lett. 75, 1260 (1995). DOI: https://doi.org/10.1103/PhysRevLett.75.1260.

\bibitem{Bousso2002}~Bousso, R. The holographic principle. Rev. Mod. Phys. 74, 825 (2002). DOI: DOI: https://doi.org/10.1103/RevModPhys.74.825.

\bibitem{Wall2009}~Wall, A. C. Ten Proofs of the Generalized Second Law. JHEP 06, 021 (2009). DOI: 10.1088/1126-6708/2009/06/021.

\bibitem{Shekh2023}~Shekh, S. H., Muzammil, M., Mapari, R. V. and Khapekar, G. U. Holographic inflation in non-static plane symmetric space-time. arXiv:2304.09080.

\bibitem{Brustein2000}~Brustein, R., Foffa, S. and Sturani, R.: Generalized second law in string cosmology. Physics Letters B, 471(4), 352-357 (2000).

\bibitem{Wang1}~ Wang, B., Abdalla, E.: Holography and generalized second law of thermodynamics in (2+1)-dimensional cosmology. Phys. Lett. B 471, 346 (2000).

\bibitem{Khadekar}~Khadekar, G.S., Kumar, P. and Islam, S.,: Modified Chaplygin gas with bulk viscous cosmology in FRW (2+ 1)-dimensional spacetime. Journal of Astrophysics and Astronomy, 40(5), 40 (2019).

\bibitem{Cornish}~Cornish, N.J., Frankel, N.E.: Gravitation in (2 + 1)-dimensions. Phys. Rev. D 43, 2555 (2000)
\bibitem{Veneziano}
Veneziano, G.: Entropy bounds and string cosmology. Phys. Lett. B 454, 22–26 (1999). arXiv:hep-th/9902126
\bibitem{Ratra} Yu, H., Ratra, B., Wang, F.-Y.: Hubble parameter and baryon acoustic oscillation measurement constraints on the Hubble constant, the deviation from spatial flatness, and other cosmological parameters. Astrophys. J. \textbf{856}, 3 (2018).

\bibitem{Moresco} Moresco, M.: Raising the bar: new constraints on the Hubble parameter with cosmic chronometers at $z \sim 2$. Mon. Not. R. Astron. Soc. \textbf{450}, L16--L20 (2015).




\end{thebibliography}
\end{document}